\documentstyle[preprint,aps]{revtex}
\begin{document}
\draft

\title{{\large\bf Effects of Next-Nearest-Neighbor Hopping on the
      Hole}\\{\large\bf Motion in an Antiferromagnetic Background}}
\author{{\bf Avraham Schiller and Pradeep Kumar}\\
        {\it National High Magnetic Field Laboratory
        and Department of Physics,\\ The University of Florida,
        Gainesville, FL 32611}\\ \vspace{10pt}
        {\bf Rainer Strack and Dieter Vollhardt}\\
        {\it Institut f\"{u}r Theoretische Physik C,\\
        Technische Hochschule Aachen, Templergraben 55\\
        D-52056 Aachen, Germany}}
\date{\today}
\maketitle

\begin{abstract}
In this paper we study the effect of next-nearest-neighbor
hopping on the dynamics of a single hole in an
antiferromagnetic  (N\'{e}el) background. In the framework
of large dimensions the Green function of a hole can be
obtained exactly. The exact density of states of a hole is
thus calculated in large dimensions and on a Bethe lattice
with large coordination number. We suggest a physically motivated
generalization to finite dimensions (e.g., 2 and 3).
In $d = 2$ we present also the momentum dependent spectral
function. With varying degree,
depending on the underlying lattice involved, the discrete
spectrum for holes is replaced by a continuum background and
a few resonances at the low energy end. The latter are
the remanents of the bound states of the $t-J$ model. Their behavior
is still largely governed by the parameters $t$ and $J$. The continuum
excitations are more sensitive to the energy scales $t$ and $t_1$.
\end{abstract}
\vspace{10pt}
\pacs{PACS numbers: $74.72.-$h, $71.27.+$a}

\section{Introduction}
The dynamics of a hole in an antiferromagnetic background is a
physically interesting but complex problem with a long history
\cite{1,nagaoka,khomskii,rice}, in particular in the theory of
magnetic semiconductors \cite{khomskii}.
In solid $^3$He \cite{he3} it precisely
corresponds to the motion of vacancies at low temperatures.
Due to the fact that in
lightly doped High-$T_{c}$ superconducting materials holes move
in the antiferromagnetic background of the original undoped
insulator \cite{6} the problem received wide attention again,
leading to exciting new physical insight.
Nevertheless a complete solution still does not exist.
The theory has recently been reviewed by Yu et al. \cite{7}.

The physical problem is most easily understood
for an Ising antiferromagnetic
(i.e. N\'{e}el) background with exchange coupling $J$
in two dimensions. As the hole moves, it breaks spin bonds.
A straight-line motion over a distance $x$ thus leads to a string of
broken bonds and thus to an increase in energy proportional to $J x$.
At the same time strings unwind themselves if the hole moves
around a loop one and a half times \cite{trugman}. Nevertheless,
 since the latter is a rather
high order process in the hopping amplitude $t$, a hole has been
considered to be bound in a linear potential
\cite{khomskii,linear}.
The resulting hole spectrum consists of a sequence of discrete
eigenvalues. For a $t - J$ model these can be seen  to scale with
$t\left(J/t\right)^{2/3}$ from
essentially dimensional arguments. Inclusion of
Heisenberg-like spin flip interactions leads in turn to
delocalization of the hole and the formation of an effective
quasiparticle band whose width is of the order of $J$. This
scenario has been studied extensively in one and two dimensions
using various analytic and numerical techniques
\cite{klr,svr,liu,dagotto,horsch,becker,schultz,7}.

We shall argue below that physical considerations suggest an
approach that involves an additional energy scale $t_{1}$, the
hopping amplitude to next-nearest-neighbor sites. A realistic band
structure simply cannot be fit by an energy dispersion that originates
solely from nearest-neighbor hopping\cite{Tremblay,maekawa,lavagna}.
In fact, estimates for $\left | t_{1}/t \right |$ range from
approximately 0.15 for ${\rm La}_{2 - x}{\rm Sr}_{x}{\rm CuO}_{4}$
to roughly 0.45 for
${\rm YBa}_{2}{\rm Cu}_{3}{\rm O}_{7-x}$\cite{Tremblay},
indicating that $t_{1}$ is far from being a small energy scale.
The significance of a $t_{1}$ term
is profound. In a bipartite lattice, motion via $t_{1}$ leaves
the spin configuration invariant. Hence the particle becomes free
to move without having to distort the spin background. While
Trugman loops \cite{trugman} can be viewed as effective $t_{1}$ processes
where $t_{1}\sim t^6$, we expect that a bare $t_{1}$ exists as an
intermediate energy scale.
We are thus led to investigate the $t-t_1-J$ model
and to ask the question \cite{20}:{\em how does $t_{1}$
alter the physical picture described above}?

For the $t-J$ model the problem of a single hole in an
antiferromagnetic background was recently solved in the limit of
large dimensions, $d \to \infty$ \cite{21}, by Metzner et al.
\cite{metzner}
for $J = 0$ and by Strack and
Vollhardt\cite{strack,24} for $J > 0$.
They were able to show that string unwindings,
whether because of Heisenberg-like spin flip interactions or
as a result of Trugman loop motion, become negligible in the
limit of infinite $d$. Hence approximations such as the Brinkman-Rice
retraceable path approximation (rpa)\cite{rice} or the linear
potential approach  \cite{khomskii,linear}
 become exact at $d = \infty$. As a result,
the density of states for a hole and its dynamical
conductivity can be calculated exactly
for $d = \infty$. Moreover, systematic improvements may be
obtained through an expansion in powers of
$1/d$\cite{metzner,strack,24}.
They show, for example, that the results of the rpa in
{\em finite} dimensions correspond to a self-consistent $1/d$-expansion
of the exact $d = \infty$ result \cite{strack,24}.
Furthermore, starting from the rpa and resumming the
Nagaoka expansion in terms of non-retraceable skeleton paths
dressed by retraceable-path insertions M\"{u}ller-Hartmann and Ventura
\cite{25} recently obtained an almost quantitative solution of the
one-particle problem in all dimensions. There is good agreement
with the numerical results by Zhong et al. \cite{26},
which were obtained for a double chain and a square
lattice with a new Lanczos iteration scheme.

For the $t-J$ model in infinite dimensions the Green function
is local and satisfies a self-consistent equation \cite{strack}
\begin{equation}
G_{ii}(z) = \langle i| \frac{1}{z - {\cal H}} | i \rangle =
\frac{1}{z - t^{\ast 2} G_{ii} \left( z - \frac{J^{\ast}}{2}
\right )} \; ,
\label{eq:gii_doo}
\end{equation}
which may be solved explicitly in terms of Bessel-functions \cite{27a}.
Here $| i\rangle$ represents the N\'eel state with
a hole at site $i$. The parameters $t^{\ast}$ and $J^{\ast}$ are
appropriately scaled  energies whose precise
definition will be given below. Eq.\ (\ref{eq:gii_doo})
coincides with  the approximation by Kane et al. \cite{klr};
the latter is therefore exact in large dimensions.

In the light of its success in capturing several of the essential
physical features  of hole motion in the $t-J$ model,
we employ the limit of large dimensions also
to the $t-t_{1}-J$ model. Exact expressions are derived
for the hole Green function, $G_{ij}(z)$, at $d = \infty$ and
a natural extension to finite dimensions is provided.
The latter continues to possess all required
analytical properties, including
spectral sum rules.

The paper is organized as follows: Sec. II contains a description
of the Hamiltonian and the necessary parameter scalings for
$d = \infty$. In sec. III, we derive the Green function for
holes on a hypercubic lattice in $d = \infty$, for a Bethe lattice
with $Z = \infty$, and for a generalization of our results
to finite $d$ ($ = 2, 3$). Sec. IV contains a discussion of our
results, followed by a summary in sec. V. In an appendix we address
the analytical properties of our finite $d$ generalization.

\section{The $t -  t_{1} - J$
model and the limit of large
dimensions}

The Hamiltonian we consider, written for convenience in terms of
creation and annihilation operators for holes, is given by
\[{\cal H} = t\sum_{\langle i,j\rangle,\sigma} n_{i,-\sigma}^{h}
h_{i\sigma}^{\dag} h_{j\sigma} n_{j,-\sigma}^{h}  +
t_{1} \sum_{\langle \langle i,j\rangle \rangle,\sigma}
n_{i,-\sigma}^{h}h_{i\sigma}^{\dag} h_{j\sigma} n_{j,-\sigma}^{h}
+ J\sum_{\langle i, j\rangle}
\vec{S}_{i} \cdot \vec{S}_{j}\]
\begin{equation}
= {\cal H}_{t} + {\cal H}_{t_{1}} +
{\cal H}_{J} \; .
\label{eq:h-ttj}
\end{equation}
Here $h_{i\sigma}$ $\left ( h_{i\sigma}^{\dag} \right )$ annihilates
(creates) a hole with spin $\sigma$ on site $i$ , and $\vec{S}_{i}$ is
the spin at site $i$. The operators $n_{i,\sigma}^{h} =
h_{i\sigma}^{\dag}h_{i\sigma}$ , which appear in the hopping terms,
ensure that no {\em electronic} double occupancy occurs in the
course of hopping. As always, $\langle i,j\rangle$ denotes
nearest-neighboring sites, while the notation
$\langle \langle i,j\rangle \rangle$
represents next-nearest-neighbors.

We adopt Eq.\ (\ref{eq:h-ttj}) as our basic model, whose microscopic
justification traces back to the Hubbard model with a
next-nearest-neighbor hopping term added.
Upon carrying out the standard canonical
transformation for a large $U$ and low hole dopings\cite{27}, a number
of terms are generated. In addition to the $J$-term $( J = 4t^{2}/U )$
specified above, we recover
an antiferromagnetic coupling $J_{1} = 4t_{1}^{2}/U$ between
next-nearest-neighbor spins, as well as a number of three-site terms. The
latter include also processes where $t$ and $t_{1}$ hoppings are combined.
Clearly, a strong enough $J_{1}$ coupling will lead to frustration
in the low temperature antiferromagnetic order at half filling. Yet, by
focusing on the physical range where $t$ constitutes an energy scale larger
than $t_{1}$ , the role of $J_{1}$ is mainly to renormalize the effective
coupling constant $J$ \cite{j_q}. We
therefore incorporate these effects in Eq.\ (\ref{eq:h-ttj})
into the (independent) parameter $J$.
 At the same time, for large $U$, $t_{1}$ itself may very well
be {\em appreciably larger} than $J$.

Figure 1 depicts the hole trajectories in the antiferromagnetic
background which become important in the presence of a
$t_{1}$ term. Here we are considering paths that bring the hole back
to its starting position without altering the spin configuration.
The most significant contribution arises from paths shown schematically in
Fig.~1(a), since they leave the spin configuration invariant. At the
same time a new type of loop motion, sketched in Fig.~1(b), appears.
Here, by combining two $t$ and one $t_{1}$ hoppings,
a triangular loop motion is sufficient to restore
the antiferromagnetic background. This must be compared with the
case for $t_{1} = 0$, where three circulations around a plaquette
are necessary for the ordinary Trugman loops.

Next we introduce the limit of large dimensions. At large $d$
the model parameters are scaled so that\cite{21,metzner,strack}
\begin{equation}
t =  \frac{t^{\ast}}{\sqrt{2d}} \;\;\; , \;\;\;
t_{1} =  \frac{t^{\ast}_{1}}{d} \;\;\; , \;\;\;
J =  \frac{J^{\ast}}{2d} \; .
\label{eq:scaling}
\end{equation}
Keeping $t^{\ast}, t_{1}^{\ast}$ and $J^{\ast}$ fixed, a meaningful
$d \rightarrow \infty$ limit emerges. One consequence of such
a limit is that spin flip processes become negligible,
so that spins may be regarded as being Ising-like \cite{metzner}.
 In addition,
without $t_{1}$, only retraceable paths are allowed up to
$O\left ( 1/d^{4}\right )$\cite{metzner}. Now, however, both
paths displayed in Fig.~1 yield non-vanishing contributions
since, for example, hopping on a triangular loop
(see Fig.~1(b) ) is proportional to $t^{2}t_{1} \sim 1/d^{2}$,
while the number of different embeddings for this path is proportional to
$d^{2}$, such that the product of the two remains of order unity even when
$d$ approaches infinity. More generally, by an appropriate combination
of $t$- and $t_1$-processes (or using only $t_1$-processes) the hole
can now move on {\em closed loops of arbitrary length}
even in $d = \infty$. Obviously the resulting motion of the
hole, consisting of simple, retraceable paths and $t_1$-generated,
background restoring loops, can be very complex.

\section{The Green function for holes}

The hole Green function is given by
\begin{equation}
G_{ij}(z) = \langle i| \frac{1}{z - {\cal H}}|j \rangle \; .
\label{eq:green}
\end{equation}
 Clearly $i$ and $j$
in Eq.\ (\ref{eq:green}) must lie on the same spin sub-lattice (taken
hereafter to be the spin up sub-lattice), otherwise $G_{ij}$ trivially
vanishes.

Typically, $G_{ij}(z)$ is calculated using Nagaoka's path
formalism\cite{nagaoka,rice}. In this formulation $\left (z -{\cal H}
\right )^{-1}$ is expanded in powers of ${\cal H}/z$, and the
different contributions are identified with all the different background
restoring paths that extend from site $j$ to site $i$. Later on we
will indeed return to this approach, but first we wish to account for
the fact that motion via $t_{1}$ alone is completely unrestricted.
For this purpose it is useful to construct two complementary
projection operators:
\begin{equation}
{\cal P} = \sum_{i \in \uparrow} |i \rangle \langle i| \;\;\;\; ,
\;\;\;\; {\cal Q} = 1 - {\cal P} \; .
\label{eq:proj}
\end{equation}
Here the summation over $i$ is confined to the spin up sub-lattice.
The hole Green function, (\ref{eq:green}), corresponds then
to the operator
\begin{equation}
\hat{G} (z) = {\cal P} \frac{1}{z - {\cal H}} {\cal P} \; .
\end{equation}
The latter may be expressed, with the aid of a familiar operator
identity\cite{foster}, as
\begin{equation}
\hat{G}(z) = \frac{1}{\hat{G}^{(0) -1}(z) - \hat{\Sigma}(z)} \; ,
\label{eq:mori}
\end{equation}
where
\begin{equation}
\hat{G}^{(0)} (z) = {\cal P} \frac{1}{z - {\cal H}_{pp}} {\cal P}
\;\;\;\; , \;\;\;\; \hat{\Sigma}(z) = {\cal H}_{pq}
\frac{1}{z - {\cal H}_{qq}}{\cal H}_{qp}
\label{eq:self}
\end{equation}
and
\begin{equation}
{\cal H}_{pp} = {\cal P} {\cal H} {\cal P}
\;\;\;\; , \;\;\;\; {\cal H}_{qq} = {\cal Q} {\cal H} {\cal Q}
\;\;\;\; , \;\;\;\; {\textstyle etc.}
\end{equation}
Here $\hat{G}^{(0)} (z)$ accounts for all pure $t_{1}$ processes,
while all processes involving $t$ are incorporated in the form
of a self-energy. This separation into spin ground-state
processes and those involving spin excitations is very useful in
simplifying the algebra. For example, ${\cal H}_{pq}$ and ${\cal H}_{qp}$
in the numerator of $\hat \Sigma(z)$
necessarily require only ${\cal H}_{t}$ and ${\cal H}_{J_{\perp}}$.
However in the denominator, ${\cal H}_{qq}$ contains all processes where
the hole moves in an excited spin state, including motion via $t_{1}$.

\subsection{The self-energy at large dimensions}
Physically, all the complexity of the hole dynamics is contained within
the self-energy. Yet at finite dimensions (i.e., $d = 2, 3$), a complete
solution for $\Sigma_{ij}(z)$ remains out of reach. We show, by contrast,
that the problem can be solved exactly at infinite dimensions,
due to a sequence of simplifications. Primarily the direct
effect of spin flips on the motion of the hole vanishes as $d
\rightarrow \infty$ \cite{metzner,strack} and
hence the self-energies can be rewritten as
\begin{equation}
\Sigma_{ij}(z) = \langle i |\hat{\Sigma}(z)| j \rangle = t^{2}
\sum_{\langle i,k \rangle} \sum_{\langle j,l \rangle}
\langle i | h^{\dag}_{i\uparrow}h_{k\uparrow}
\frac{1}{z - {\cal Q} \left ( {\cal H}_{t} + {\cal H}_{t_{1}} +
{\cal H}_{J_{z}} \right) {\cal Q}} h^{\dag}_{l\uparrow}h_{j\uparrow}
| j \rangle \; .
\label{eq:self_z}
\end{equation}

Next, as usual in  $d = \infty$, $\Sigma_{ij}(z)$ becomes
site-diagonal \cite{21}.
This characteristic large dimensions feature follows
from the application of Nagaoka's path formalism to
(\ref{eq:self_z}). A typical path describing $\Sigma_{ij}(z)$ begins
with an excitation of the ground-state spin configuration at site $j$, and
its restoration only at the final step when the hole reaches site $i$.
In between, the spin background is continuously excited during the entire
hole motion. Consider now the case where $i$ and $j$ are two different
sites (Fig.~2). Then the fact that the spin background is eventually restored
at a site $i$ different than $j$ demands that the latter must be revisited
by the traveling hole at some intermediate step. Otherwise, the down
spin it acquired in the initial step will never be mended, as it should, to
an up spin. Since the overall spin background remains
excited through all intermediate stages, this implies that some other
site $f$ on the up sub-lattice is forced to acquire a down spin,
in between the two hole visits to site $j$. Moreover, after the spin at
site $j$ is finally corrected to an up one, site $f$ has
to be revisited in the course of restoring the ground-state spin
configuration. A schematic description of the path is shown
in Fig. 2. It consists, among other things, of at least
three segments where the hole propagates between the initial site $j$
and the intermediate site $f$. From the standpoint of large dimensions
this is two segments too many. Thus the effect of the spin
background on the off-diagonal components of the self-energy is to
reduce their contribution  by an integer power of $1/d^{2}$. Indeed
the path illustrated in Fig.~2 does not have the topology
of a ``loop tree'' \cite{metzner}
and hence does not contribute at $d = \infty$.

The third and final simplification at large dimensions occurs
in the calculation of the remaining on-site self-energy. In
order to evaluate $\Sigma_{ii}(z)$ we focus on the following
functions
\begin{equation}
\Sigma_{ii}^{(k,l)}(z) =
\langle i | h^{\dag}_{i\uparrow}h_{k\uparrow}
\frac{1}{z - {\cal Q} \left ( {\cal H}_{t} + {\cal H}_{t_{1}} +
{\cal H}_{J_{z}} \right) {\cal Q}} h^{\dag}_{l\uparrow}h_{j\uparrow}
| i \rangle
\label{eq:self_k_l}
\end{equation}
($k, l$ being nearest neighbors of $i$), and compare them with
\begin{equation}
G_{kl}(z) = \langle {\rm N\acute{e}el} |  h^{\dag}_{k\downarrow}
\frac{1}{z - {\cal H}_{t} + {\cal H}_{t_{1}}}
h_{l\downarrow} | {\rm N\acute{e}el} \rangle \; .
\label{eq:g_k_l}
\end{equation}
Here $| {\rm N\acute{e}el} \rangle$ designates the antiferromagnetic spin
ground-state with no holes present. Eq.\ (\ref{eq:g_k_l}) is
no other than the hole Green function, only written in terms of
a hole in the down spin sub-lattice rather than the up spin one.
$\Sigma_{ii}^{(k,l)}(z)$ and $G_{kl}(z)$ differ in two respects:
(1) the ${\cal Q}$ projection operator appears in the denominator
of Eq.\ (\ref{eq:self_k_l}), and (2) the spin at
site $i$ is flipped in Eq.\ (\ref{eq:self_k_l}) to a down spin.
Both differences are conveniently accounted for in the limit of
large dimensions by resorting once again to Nagaoka's path formalism,
and pursuing the same line of reasoning as before. Here one can see
that all possible hole trajectories where site $i$ is involved are
reduced by an integer power of $1/d$ for both
$\Sigma_{ii}^{(k,l)}(z)$ and $G_{kl}(z)$.
Hence the spin at site $i$ may be regarded as being {\em frozen} in
each of the two cases. Such an observation yields profound
consequences. To begin with, the down spin at site $i$ in the case of
$\Sigma_{ii}^{(k,l)}(z)$ ensures that the spin background remains
constantly excited. Therefore the ${\cal Q}$ projection operators
entering Eq.\ (\ref{eq:self_k_l}) are automatically satisfied.
In addition, since apart from the flipped spin at site $i$ the
initial and final spin configurations in Eq.\ (\ref{eq:self_k_l})
are identical to those of Eq.\ (\ref{eq:g_k_l}), we precisely recover
the same path contributions for both cases. Thus the {\em only}
distinction between the two cases exists in their Ising energy,
which happens to be $J^{\ast}/2$ larger for $\Sigma_{ii}^{(k,l)}(z)$.
This is a result of the broken bonds between the flipped spin
at site $i$ and its surrounding nearest neighbor spins. All in all we
arrive at the following relation, which becomes exact for large
dimensions:
\begin{equation}
 \Sigma_{ii}^{(k,l)}(z)
= G_{kl}(z - \frac{J^{\ast}}{2}) \; .
\label{eq:relation}
\end{equation}

Finally, the combination of Eqs.\ (\ref{eq:self_z}) and
(\ref{eq:relation}) yields
\begin{equation}
 \Sigma (z) \equiv \Sigma_{ii}(z)  = t^{2}
\sum_{\langle i,k \rangle} \sum_{\langle i,l \rangle}
G_{kl}(z - \frac{J^{\ast}}{2}) \; .
\label{eq:self_doo}
\end{equation}

Note that no restrictions on the nature of the underlying lattice
were involved in the derivation of equation (\ref{eq:self_doo}),
hence it equally applies to {\em all} bipartite lattices in the
limit of infinite number of nearest neighbors.

\subsection{The hole Green function at large dimensions -
hypercubic lattice}
After establishing that the self-energy is site-diagonal,
Eq.\ (\ref{eq:mori}) is readily solved for a hypercubic
lattice by Fourier transformation:
\begin{equation}
G_{\vec{k}}(z) = \frac{1}{z - \epsilon_{\vec{k}} -
\Sigma(z)} \; ,
\label{eq:g_of_k}
\end{equation}
with
\begin{equation}
\epsilon_{\vec{k}} = 2 t_{1} \sum_{n \neq m}
\cos (k_{n})\cos (k_{m}) \; .
\label{eq:baredos}
\end{equation}
Here $\vec{k}$ lies within a reduced Brillouin zone, and the lattice
constant is taken to be unity. The spatial Green functions are obtained
from the inverse transform of Eq.\ (\ref{eq:g_of_k}). It yields
\begin{equation}
G_{ij}(z) = G_{ij}^{(0)}(z - \Sigma(z)) \; ,
\label{eq:g_ij}
\end{equation}
where $G_{ij}^{(0)}$ is the free, tight-binding Green function due to
next-nearest-neighbor hopping. Next we make use of the cubic symmetry,
and express $\Sigma(z)$ from Eq.\ (\ref{eq:self_doo}) as
\begin{equation}
\Sigma(z) = t^{\ast 2} \left \{ G(z - \frac{J^{\ast}}{2})
+ 2(d - 1) G_{1}(z - \frac{J^{\ast}}{2}) + G_{2}(z - \frac{J^{\ast}}{2})
\right \} \; .
\label{eq:self_any_d}
\end{equation}
Here the notation $G \equiv G_{ii}$
and $G_{1}$ was introduced for the on-site
and next-nearest-neighbor Green functions, respectively, while $G_{2}$
corresponds to the case where the two lattice sites lie on the same
axis, yet with a separation of two lattice constants (i.e.,
third-order-nearest neighbors). The former two functions are further related
through an identity that links the corresponding local and next nearest
neighbor tight binding Green functions. Indeed for any dimension
\begin{equation}
2(d - 1) G^{(0)}_{1}(z) =
\frac{z G^{(0)}(z) - 1}{t_{1}^{\ast}} \; .
\label{eq:g1_via_g0}
\end{equation}
where $G_{ii}^{(0)} \equiv G^{(0)}$.
Thus, since $G_{2}(z)$ vanishes at large dimensions as $1/d$,
Eqs.\ (\ref{eq:self_any_d}) and (\ref{eq:g1_via_g0}) reduce at
the limit of $d \rightarrow \infty$ to
\begin{equation}
\Sigma(z) = t^{\ast 2} \left \{ G(z - \frac{J^{\ast}}{2})
+ \frac{(z - \frac{J^{\ast}}{2} - \Sigma(z - \frac{J^{\ast}}{2}) )
G(z - \frac{J^{\ast}}{2}) - 1}{t_{1}^{\ast}} \right \} \; .
\label{eq:self_final}
\end{equation}

Together, equations (\ref{eq:g_ij}) and (\ref{eq:self_final}) combine
to give an iterative equation for the local Green function
$G(z)$, which is exact at $d = \infty$. It must be supplemented,
however, with the corresponding $d = \infty$ expression for
$G^{(0)}(z)$ which
may be obtained (using the corresponding density of
states \cite{muller}) as \cite{lech}
\begin{equation}
G^{(0)} (z) =
\frac{-i}{ 2t^{\ast}_{1} }
\frac{\sqrt{\pi}}{ \sqrt{(z + t^{\ast}_{1})/2t^{\ast}_{1}} }
W \left( \sqrt{ \frac{z + t^{\ast}_{1}}{2t^{\ast}_{1}} }
\right ) \; .
\label{eq:dos_hyp}
\end{equation}
Here $W(z)$ is the scaled complementary error function of a complex
argument\cite{stegun}, and the sign of the square root is chosen so
that it lies in the upper half plane.

\subsection{The hole Green function on a Bethe lattice}
In case of a Bethe lattice, the definitions of the scaled parameters
are modified as
\begin{equation}
t =  \frac{t^{\ast}}{\sqrt{Z}} \;\;\; , \;\;\;
t_{1} =  \frac{t^{\ast}_{1}}{Z} \;\;\; , \;\;\;
J =  \frac{J^{\ast}}{Z} \; ,
\label{eq:scaling_bethe}
\end{equation}
where $Z$ is the coordination number.
Once these adjustments are carried out, equations (\ref{eq:g_ij})
and (\ref{eq:self_final}) continue to hold as the exact $Z = \infty$
equations for the hole Green function $G_{ij}(z)$
where now, however, $G^{(0)}(z)$ in Eq.\ (\ref{eq:self_final})
 is given by \cite{rice}
\begin{equation}
G^{(0)}(z) = \frac{1 - \sqrt{ 1 - 4t^{\ast}_{1}/
(z + t^{\ast}_{1})} }{2t^{\ast}_{1}} \; .
\label{eq:dos_bethe}
\end{equation}

\subsection{Extension to finite dimensions}
A systematic $1/d$ expansion about the $d = \infty$ limit constitutes
in general a formidable task, and in that respect the problem of hole
motion in the $t-t_{1}-J$ model is no exception. Here we consider a
hypercubic lattice and suggest an alternative, approximate way to
bridge between $d = \infty$ and  finite $d$.

We base our approach on the same equations that were derived for large
dimensions, yet by two adjustments. First, the $G_{ij}^{(0)}(z)$
Green functions are replaced by their corresponding $d$-dimensional
expressions. Second, we retain the contribution of $G_{2}$ to the
local self-energy, i.e. use Eq.\ (\ref{eq:self_any_d}) rather than
Eq.\ (\ref{eq:self_final}). Though the latter contribution completely
vanishes as $d \rightarrow \infty$, it is essential to keep it
at finite $d$ in order to avoid a non-physical spectral function
(note that for $d < \infty$
 a shift occurs in the band edge of the
spectral part of $G^{(0)}(\omega - i\delta)$;
see, e.g., Fig. 3). We elaborate
on this point in the appendix. The complete set of equations
employed for a finite $d$ consists, therefore,
of Eq.\ (\ref{eq:g_ij}), together with the following version of
Eq.\ (\ref{eq:self_any_d}):
\begin{equation}
\Sigma(z) = t^{\ast 2} \left \{ G(z - \frac{J^{\ast}}{2})
+ \frac{(z - \frac{J^{\ast}}{2} - \Sigma(z - \frac{J^{\ast}}{2}) )
G(z - \frac{J^{\ast}}{2}) - 1}{t_{1}^{\ast}}  +
G_{2}(z - \frac{J^{\ast}}{2})\right \} \; .
\label{eq:self_finite_d}
\end{equation}

We further note that the only input function required at $d = \infty$
is $G^{(0)}(z)$, while for a finite $d$ it is joined by
$G_{2}^{(0)}(z)$. For $d = 2, 3$, both functions can be computed using
complete elliptic integrals of the first and second kind, provided
they are properly continued to the entire complex
plane\cite{fcc}.
In the appendix we show that all analytical properties required
from $G_{ij}(z)$ are indeed conserved within the approximation scheme
suggested above.

\section{Numerical Results}

\subsection{The bare density of states}
Figure 3 shows the bare density of states
(DOS) due to pure $t_1$-hopping on a hypercubic lattice
in $d = 2$, corresponding to the DOS for a square lattice, in
$d = 3$ and $d = \infty$ [see Eq. (21)], corresponding
to the DOS for a fcc-lattice in those dimensions, and, finally,
for a Bethe lattice with $Z = \infty$, Eq.~(23).
 There is a divergence at the lower band edge (for
$t_{1}^{\ast} > 0$), either with a square root singularity
for $d = \infty$ and the Bethe lattice, or in a logarithmic
manner\cite{fcc} for $d = 3$. For $d = 2$ the logarithmic
singularity is at the band center. The curves for $d = \infty$
and Bethe lattice are clearly very similar, although the bandwidth
for a Bethe lattice remains finite. In contrast, $d = \infty$
possesses an exponential tail that extends to higher energies
\cite{muller}.
For $d = 2$ and $3$ there is a slight shift in the lower
band edge as mentioned earlier in sec. III.D.
Except for $d = 2$ the cases look
qualitatively similar. For $t_{1}^{\ast} < 0$ the bare
density of states are inverted (i.e., $\omega \rightarrow
-\omega$). The divergences then move to the upper band edge,
with the exception of the $d = 2$ case which is symmetric
in $\omega$.

\subsection{Infinite dimensions}

Fig.~4 shows the effect of an increase in $t_1^{\ast} > 0$ on the DOS
for the motion of a hole in the infinite-dimensional $t-t_1-J$ model
at fixed $J^{\ast}/t^{\ast} = 0.4$.
For $t_{1}^{\ast}/t^{\ast} = 0.05$ we see only little changes from
the case where $t_{1}^{\ast} = 0$: there are new satellites
in-between the original bound state energies, starting at
$\omega_{0} + J^{\ast}/2$, where $\omega_{0}$ is the lowest
bound state energy; the sharp delta peaks
for $t_{1}^{\ast} = 0$ have now acquired small width.
(Note, however, that we expect the realistic energy hierarchy to be
$t^{\ast} > t_{1}^{\ast} \agt J^{\ast}$.)
At $t_{1}^{\ast}/t^{\ast} = 0.2$ a
continuous spectrum has developed. The sharp
structures near the lower band edge are derived from the
divergence of the bare DOS. At
$t_{1}^{\ast}/t^{\ast} = 0.4$ only the lowest resonances
appear discernible. The rest has been absorbed into the incoherent
continuum of excitations. Any further increase in $t_{1}^{\ast}$
leads to additional decrease of the resonant peaks
and a further development of the underlying continuum. It looks
as if one is approaching the general shape of the bare
(continuous) density of states.

For $t_{1}^{\ast} < 0$, Fig. 5, there is a dramatic change
in the DOS. By $t_{1}^{\ast}/t^{\ast} = -0.2$ only weak traces
are left of the  bound states at $t_{1}^{\ast} = 0$. The DOS
is more symmetric and less structured compared to
$t_{1}^{\ast} > 0$. It corresponds to an incoherent band with
an exponentially small tail that extends to negative energies.
The origin of the strong effect that a negative $t_{1}^{\ast}$
has on the hole motion lies in the unboundedness of the
bare DOS for $\omega \rightarrow -\infty$.
As a result, the self-energy acquires a nonzero imaginary
part throughout the negative energy region.

Figure 6 displays the DOS for a Bethe lattice with $Z = \infty$
for $J^{\ast}/t^{\ast} = 0.4$ and $t_{1}^{\ast}/t^{\ast} =
\pm 0.4$. Again, well-developed incoherent continua are
recovered, similar to the ones
obtained for $d = \infty$. The resonances persist, however,
even for $t_{1}^{\ast}/t^{\ast} = -0.4$. They seem to be the
dominant structures for smaller values of $|t_{1}^{\ast}|$
($|t_{1}^{\ast}| \alt 0.2$).
Notice also the shift in the threshold energy between
$t_{1}^{\ast} > 0$ and $t_{1}^{\ast} < 0$. For $t_{1}^{\ast} < 0$
the resonances shift to lower energies with increasing
$|t_{1}^{\ast}|$, while in case of
$t_{1}^{\ast} > 0$ the lower edge is quite insensitive to
variations in $t_{1}^{\ast}$. The difference between
$d = \infty$ and the Bethe lattice is mostly due to the
finite width of the bare spectrum of the latter.

\subsection{Finite dimensions}
Next we turn to finite dimensions.
Figure 7 depicts our generalization to $d = 3$  for
$J^{\ast}/t^{\ast} = 0.4$ and $t_{1}^{\ast}/t^{\ast} =
\pm0.2, \pm 0.4$. The width and magnitude of the
incoherent spectra seem to be similar for $t_{1}^{\ast} < 0$
and $t_{1}^{\ast} > 0$, as for all lattices discussed here.
The lower edge is found, once again, to have no significant
shift with $t_{1}^{\ast}$ for $t_{1}^{\ast} > 0$,
while it decreases with decreasing $t_{1}^{\ast}$ for
$t_{1}^{\ast} < 0$. The qualitative behavior of the DOS for
$t_1^{\ast}/t^{\ast} = 0.4$ is similar
to that obtained by Bala et al. \cite{20} for $d = 2$ using a very
different method. The effect of a smaller $J^{\ast}$,
corresponding to what might be expected for realistic material
parameters, is shown in  Fig. 8;
for $J^{\ast}/t^{\ast} = 0.1$
and $t_{1}^{\ast}/t^{\ast} = -0.4$, we recover a broad
incoherent spectrum.
At the same time, resonances can be clearly seen at the
lower edge, with widths that are smaller than the separation
between adjacent peaks. The low-energy features thus persist
even for $J^{\ast}$ appreciably smaller than $t_{1}^{\ast}$.

We now examine more closely the low-energy spectrum for two
dimensions.
For $d = 2$, the general features described above
are seen as well. The effect of increasing
$|t_{1}^{\ast}|$ is to broaden the original bound states into
resonances, eventually leading to continuum in
the DOS. There are, however, two major differences.
The first is that a larger $|t_{1}^{\ast}|$ is needed to
produce qualitatively similar results in $d = 2$, as
compared to $d = 3$. The reason is simple: the overall
bandwidth of the bare DOS is smaller for $d = 2$
(see Fig. 3). The other difference is that in $d = 3$ a
negative $t_{1}^{\ast}$ is seen to be more effective in smearing
out the residual details of the bound states. In $d = 2$, the
roles are slightly inverted. This has to do with the symmetric
character of the bare DOS in $d = 2$ as opposed to
the strong asymmetry for $d \ge 3$.

We have also studied the $\vec{k}$-dependent spectral function
$A(\vec{k},\omega)$ for $d = 2$. Figure 9 shows
$A(\vec{k},\omega)$ for $\vec{k} = (\frac{\pi}{2},
\frac{\pi}{2})$ and $J^{\ast}/t^{\ast} = 0.2$,
$t_{1}^{\ast}/t^{\ast} = -0.6$. It looks quite similar to the
spectral function calculated numerically by Liu and
Manousakis\cite{liu}  for the $t-J$ model on a finite size
lattice ($J^{\ast}/t^{\ast} = 0.2$ corresponds, for $d = 2$,
to $J/t = 0.1$). The lowest peak in this curve corresponds to
a coherent state. Indeed {\em all} $\vec{k}$-states within a
margin of $J^{\ast}/2$ from the lower band edge have vanishing
imaginary part of the self-energy.
Physically, this corresponds to their inability to decay by
emitting or absorbing spin excitations, which is a
consequence of the Ising nature of the spins. For
$t_{1}^{\ast} < 0$ the dispersion of this coherent band is
similar to that of the bare energy in
Eq.\ (\ref{eq:baredos}), with a strong renormalization due
to $J^{\ast}$. To be more specific, we find that
\begin{equation}
E(\vec{k}) = \frac{J^{\ast}}{2}f \left(\vec{k};
t_{1}^{\ast} \right ) + constant \; .
\end{equation}
For a fixed $t_{1}^{\ast}$ and different $J^{\ast}$,
the eigenvalues scale with $J^{\ast}/2$ in that
$ \left [ E(\vec{k}) - E(\vec{k}_{0}) \right ]/
\left ( \frac{J^{\ast}}{2} \right )$
all lie on the same curve. This is demonstrated in Fig. 10(a).
The function $f$, however, differs from Eq.\ (\ref{eq:baredos})
in that the upper band edge is deformed towards lower energies
for nonnegligible values of $t_{1}^{\ast}$ [see Fig. 10(b)].
For small values of $t_{1}^{\ast}$, though, $f$ is essentially
linear in $t_{1}^{\ast}$.
The overall integrated spectral weight of the entire band can
be expressed as:
\begin{equation}
\int d\epsilon_{\vec{k}} \rho^{(0)}(\epsilon_{\vec{k}})
\frac{d E({\vec{k}})}{d \epsilon_{\vec{k}}} \; ,
\label{eq:z_weight}
\end{equation}
where $\rho^{(0)}$ is the bare DOS for $d = 2$.
Following $E({\vec{k}})$, Eq.\ (\ref{eq:z_weight}) also
scales with $J^{\ast}$.
One consequence of the self-energy being $\vec{k}$-independent
in our approximation is that the minimum of $E(\vec{k})$
always occurs at the $\Gamma$ point, $\vec{k} = (0,0)$.

For positive $t_{1}^{\ast}$ in $d = 2$, one notable feature is
that for moderate $t_{1}^{\ast}$
($t_{1}^{\ast} \alt J^{\ast}$), when the low energy resonance
in the DOS has almost become part of the continuum,
there are wave vectors for which coherent state solutions no longer
exist. Three types of solutions are then recovered. At low
frequencies, and over a frequency range of
$\Delta \omega = J^{\ast}/2$, true coherent state
solutions are found. These are centered around the band minima
at $\vec{k} = (\pi,0)$.
At slightly higher frequencies, the imaginary part
of the self-energy becomes nonzero in a quadratic manner.
The solutions are thus quasiparticle-like. In the
neighborhood of $\vec{k} = (0,0)$ --- the upper band edge ---
and over an area in $\vec{k}$-space that increases with
increasing $t_{1}^{\ast}$, the solutions are removed
from the low energy side and are shifted to the incoherent
high frequency side of the spectrum.

\section{Discussion}
In this paper we explored the consequences of a next-nearest-neighbor
 hopping $t_{1}$ on the motion of a single hole
in a quantum antiferromagnet. The system is described by a $t-t_{1}-J$
model. Our motivation stems both from the need of a $t_{1}$
term to characterize the observed electronic
structure\cite{Tremblay,maekawa,lavagna}, and from
the growing numerical evidence\cite{dagotto-prl}  that the
effective quasiparticle dispersion generated by the
$t-J$ model can be parameterized by means of effective
next-nearest-neighbor and third-order-nearest-neighbor hops.
Our approach is based on the limit of large dimensions,
where the problem is solved exactly. For $d = \infty$,
$t_{1}$ survives as the {\em only} mechanism that delocalizes
the hole. The principal effect of the introduction of
$t_{1}$ is thus seen in the following steps:
(a) For $t_{1}^{\ast} = 0$ (and $J^{\ast} \neq 0$), we
have only bound states that are typically separated by
$t(J/t)^{2/3}$; (b) as $t_{1}^{\ast}$ increases
($t_{1}^{\ast} \ll J^{\ast}$), all peaks in the density of
states acquire width, and satellites appear at intermediate
energies $\epsilon_{n,m} =
\omega_{n} + a_{n,m}J^{\ast}/2$ (here $\omega_{n}$ is the
energy of the $n$-th bound state and $a_{n,m} \sim O(1)$);
(c) as $t_{1}^{\ast}$ further increases ($t_{1}^{\ast} \sim
J^{\ast}$), the satellites and resonant peaks merge, and
a continuum is formed. Hence $t_{1}$ restores the incoherent
Brinkman-Rice\cite{rice} spectra at high energies,
while generating a quasiparticle dispersion at low energies.
This is most clearly seen within our analysis in $d = 2$.
For $|t_{1}^{\ast}| \ll J^{\ast}$
and $\omega$ close to the low-energy edge, we may
replace the self-energy in Eq.\ (\ref{eq:g_ij}) with its
$t_{1}^{\ast} = 0$ limit. Then, using a dominant pole
approximation\cite{klr} ,
\begin{equation}
G_{\vec{k}}(\omega + i\delta) \cong \frac{Z_{0}}{\omega -
\omega_{0} - Z_{0}\epsilon_{\vec{k}}} \; ,
\end{equation}
where $\omega_{0}$ and $Z_{0}$ are the $t_{1}^{\ast} = 0$
ground-state energy and quasiparticle amplitude, respectively.
Hence a quasiparticle band immediately appears, with a characteristic
renormalization $Z_{0}$. Its bandwidth is given by $4Z_{0}
t_{1}^{\ast}$ (the factor 4 is for $d = 2$), while the
$\vec{k}$-independent effective mass is proportional to
$1/\left ( Z_{0}t_{1}^{\ast} \right )$. The effect of
$J^{\ast}$ enters via the quasiparticle amplitude $Z_{0}$.
In the intermediate and large regimes for $t_{1}^{\ast}/J^{\ast}$,
the bandwidth is practically governed by $J^{\ast}$
(see, e.g., Fig. 10). For $t_{1}^{\ast} < 0$, the entire
Brillouin Zone is included in this band, and the quasiparticle
dispersion scales with $J^{\ast}$. At moderate
values of $t_{1}^{\ast} > 0$, quasiparticles solutions are
removed from the vicinity of the $\Gamma$ point.
In both cases, the frequency range of width $J^{\ast}/2$ at the
lowest band edge accommodates only coherent states. This
corresponds to the inability of holes to loose energy by
flipping the spins. The latter is a
consequence of large dimensions
where the original Heisenberg spins become
Ising-like, i.e. quantum spin fluctuations are suppressed.
Quantum fluctuations are known
to have two important effects in the $t-J$ model. Together
with Trugman loops\cite{trugman}  they act to delocalize the
hole, giving rise to an effective quasiparticle energy
dispersion that can be fitted by effective hops to
next-nearest-neighbor and third-order-nearest-neighbor sites.
Quantum fluctuations are also responsible for
the quasiparticle-band minimum at
$\vec{k} = (\frac{\pi}{2}, \frac{\pi}{2})$ (see, e.g., Zhou
and Schultz\cite{schultz} for a clear demonstration of this
point). The  introduction of a $t_{1}$-term to the  model, motivated
by physical considerations,  simulates quite well the
dispersion, i.e. delocalization,
of the hole but  cannot produce a band
minimum at $\vec{k} = (\frac{\pi}{2}, \frac{\pi}{2})$.

In summary, we have provided a simple analytic theory for the
hole-motion in an antiferromagnetic background in the presence of
next-nearest-neighbor hopping. The theory is exact in the
limit of large dimensions. In finite dimensions it
 provides a useful mean-field
description of this physically relevant phenomenon.

Two of us (A.S. and P.K.) are grateful to L. Borkowski for stimulating
discussions at early stages of this work.
D.V. acknowledges a useful discussion with P.Horsch. This
work was supported by the NHMFL (A.S. and P.K.),
DOE DE-FG05-91ER45462 (P.K.) and the Deutsche Forschungsgmeinschaft
through Sonderforschungsbereich 341 (R.S. and D.V.).

\begin{appendix}
\section*{}
In this appendix we show that our generalization to finite
dimensions, Eq.\ (\ref{eq:self_finite_d}), retains all
analytical properties in the upper and lower half planes,
as required of $G_{ij}(z)$. Our strategy is as follows:
First we show that the imaginary parts of $\Sigma(z)$
and $z$ have opposite signs, which implies that $z$ and
$z - \Sigma(z)$ always lie in the same half plane.
In particular, $z - \Sigma(z)$ has a nonvanishing
imaginary part so long as $z$ is not purely real.
Then, making use of the known analytical structure of
$G_{ij}^{(0)}(z)$ we conclude that: (1) $G_{ij}(z) =
G_{ij}^{(0)}(z - \Sigma(z))$ has no poles in either
the upper or lower half planes; (2) $\Im m G(z)$
has a fixed sign throughout each of the two half planes.
The latter point guarantees that the spectral part of
$G$ is non-negative. Finally, the spectral sum rule
(i.e., the normalization of the spectral function to unity)
is a direct consequence of the fact that $G(z)$ falls
like $1/z$ for large $|z|$ and has no poles outside the
real axis.

We now construct the first and principal step in the
procedure outlined above. From the asymptotic behavior of
$G_{ij}^{(0)}(z)$ it follows that $G(z) =
G^{(0)}(z - \Sigma(z))$ falls like $1/z$ for large
$|z|$, while $G_{1}(z)$ and $G_{2}(z)$ decay according to
higher powers of $1/z$. Thus by virtue of
Eq.\ (\ref{eq:self_any_d}), $\Sigma(z) \sim
\frac{t^{\ast 2}}{z}$ for $|z| \rightarrow \infty$.
Asymptotically it is therefore apparent that the imaginary
parts of $\Sigma(z)$ and $z$ acquire opposite signs.
This relation can now be extended to the entire upper and
lower half planes using the recursion relation,
Eq.\ (\ref{eq:self_any_d}). Replacing $G(z), G_{1}(z)$
and $G_{2}(z)$ with explicit $k$-space integrations,
Eq.\ (\ref{eq:self_any_d}) is rewritten as
\begin{equation}
\Sigma(z + \frac{J^{\ast}}{2}) =
\int_{-\pi}^{\pi} \prod_{l = 1}^{d} \frac{dk_{l}}{2\pi} \;
\frac{t^{\ast 2}}{ z - \Sigma(z) - \epsilon_{\vec{k}} }
\left \{ 1 + \frac{2}{d}\sum_{n \neq m}\cos (k_{n})\cos (k_{m})
+ \frac{1}{d}\sum_{n = 1}^{d} \cos (2k_{n}) \right \} \;.
\label{eq:app1}
\end{equation}
Here the different terms in the curly brackets correspond to
$G(z), G_{1}(z)$ and $G_{2}(z)$, respectively. These may
be easily brought to the form
\begin{equation}
\Sigma(z + \frac{J^{\ast}}{2}) =\frac{2t^{\ast 2}}{d}
\int_{-\pi}^{\pi} \prod_{l = 1}^{d} \frac{dk_{l}}{2\pi} \;
\frac{\left ( \sum_{n = 1}^{d} \cos (k_{n}) \right )^{2}}
{ z - \Sigma(z) - \epsilon_{\vec{k}} } \;,
\label{eq:app2}
\end{equation}
which clarifies the positivity of the numerator in the integrand.
Explicitly taking the imaginary part of the right-hand side in
Eq.\ (\ref{eq:app2}), we find that
$\Im m \Sigma(z + \frac{J^{\ast}}{2})$ and $\Im m \left \{
z - \Sigma(z) \right \}$ always have opposite signs. Thus,
given that asymptotically $\Im m \Sigma(z)$ is negative
(positive) in the upper (lower) half plane, and since any point
in the half plane can be approached from the asymptotic regime
by way of repeated applications of the recursion relation,
$\Im m \Sigma(z)$ remains negative (positive) throughout
the {\em entire} half plane. Hence the imaginary parts
of $z$ and $\Sigma(z)$ always acquire opposite signs.

Having established this point, the analytical properties of
$G_{ij}(z)$ follow along the lines illustrated above.
Both the positivity of the spectral function and the spectral
sum rule have also been confirmed numerically for several
cases, including those displayed in Figs. 7, 8 and 9.

Finally, we wish to elucidate the importance of retaining
$G_{2}(z)$ in Eq.\ (\ref{eq:self_finite_d}). If $G_{2}$ is
eliminated from Eq.\ (\ref{eq:self_finite_d}) we violate,
for any finite $d$, the positivity of the curly brackets
in Eq.\ (\ref{eq:app1}). Consequently
$\Im m \Sigma(z)$ changes signs both within the upper
and lower half planes, as does the imaginary part of
$z - \Sigma(z)$. The branch cut $G_{ij}^{(0)}(z)$
possesses on the real axis thus shifts into the upper and
lower half planes, and $G_{ij}(z)$ is no longer analytic.
\end{appendix}

\begin{figure}
\caption{Hole trajectories that become important in the presence of
a $t_{1}$ term. (a) Diagonal motion; (b) Triangular loop motion.}
\end{figure}

\begin{figure}
\caption{A schematic description of the paths that contribute
to the off-diagonal self-energy, $\Sigma_{ij}$. The hole
starts its motion at site
$j$. It approachs the intermediate site $f$ for the first time
(segment 1) before revisiting site $j$ (segment 2). Then it
completes its motion to site $i$ by way of $f$
(segments 3 and 4).}
\end{figure}

\begin{figure}
\caption{Bare DOS due to pure $t_1$-hopping on a hypercubic
lattice in $d = 2, 3, \infty$, and on a Bethe lattice
with an infinite coordination number, $Z = \infty$.}
\end{figure}

\begin{figure}
\caption{DOS for $J^{\ast}/t^{\ast} = 0.4$
and $t^{\ast}_{1}/t^{\ast} = 0.05,
0.2$ and $0.4$ for a hypercubic lattice at $d = \infty$.}
\end{figure}

\begin{figure}
\caption{Similar to Fig. 4, but with $t^{\ast}_{1}/t^{\ast} = -0.1$
and $-0.2$. Notice the disappearance of all discrete (resonant)
structures for $t^{\ast}_{1}/t^{\ast} = -0.2$ .}
\end{figure}

\begin{figure}
\caption{Bethe lattice with coordination number $Z = \infty$ :
DOS for $J^{\ast}/t^{\ast} = 0.4$ and
$t^{\ast}_{1}/t^{\ast} = \pm 0.4$.}
\end{figure}

\begin{figure}
\caption{DOS in $d = 3$ [see
Eq.\ (\protect\ref{eq:self_finite_d})] : DOS
for $J^{\ast}/t^{\ast} = 0.4$ and
$t^{\ast}_{1}/t^{\ast} = \pm 0.2, \pm 0.4$ .}
\end{figure}

\begin{figure}
\caption{Same as figure 7 with $J^{\ast}/t^{\ast} = 0.1$ and
$t^{\ast}_{1}/t^{\ast} = -0.4$ .}
\end{figure}

\begin{figure}
\caption{The $A(\vec{k},\omega)$ spectral function for $d = 2$
and $\vec{k} = (\frac{\pi}{2}, \frac{\pi}{2})$. Here
$J^{\ast}/t^{\ast} = 0.2$ and $t_{1}^{\ast}/t^{\ast} = -0.6$ .}
\end{figure}

\begin{figure}
\caption{The quasiparticle dispersion, $E(\vec{k})$,
for $d = 2$, (a) as a function of $J^{\ast}$ for
fixed $t_{1}^{\ast}/t^{\ast} = -0.6$, (b)
as a function of $t_{1}^{\ast}$ for fixed
$J^{\ast}/t^{\ast} = 0.1$. Here $\Gamma = (0,0)$,
$M = (\pi,0)$ and $X = (\frac{\pi}{2},\frac{\pi}{2})$ .}

\end{figure}


\begin{references}
\bibitem{27a} We note that the result for $G_{ii}$
by M. M. Mohan [J. Phys.: Condens. Matter {\bf 3},
4507 (1991)] derived on the basis of the rpa is correct even to order $1/d$
 and hence coincides with the solution of Eq.(1) at $d = \infty$.
\bibitem{j_q} For classical spins, zero temperature antiferromagnetic order
will persist at half filling and $d \rightarrow \infty$ throughout
the range $t_{1}^{\ast 2} < t^{\ast 2}/2 $. Here $t_{1}^{\ast}$ and
$t^{\ast}$ are the properly scaled hopping matrix elements (see Eq.
(\ref{eq:scaling})).
\end{references}
\end{document}